
\input phyzzx.tex

\def\a{\alpha}
\def\b{\beta}

\def\G{\Gamma}
\def\d{\delta}
\def\D{\Delta}

\def\z{\zeta}
\def\t{\theta}

\def\r{\rho}

\def\k{\kappa}
\def\l{\lambda}

\def\o{\omega}

\def\s{\sigma}

\def\pa{\partial}

\def\ov{\overline}
%
\def\ap#1{{\it Ann. Phys.} {\bf #1}}
\def\cmp#1{{\it Comm. Math. Phys.} {\bf #1}}

\def\pl#1{{\it Phys. Lett.} {\bf B#1}}
\def\prl#1{{\it Phys. Rev. Lett.} {\bf #1}}
\def\prd#1{{\it Phys. Rev.} {\bf D#1}}

\def\np#1{{\it Nucl. Phys.} {\bf B#1}}

%
\REF\hawk{S. W. Hawking, \cmp{43} (1975) 199.}
\REF\call{C.G. Callan, S.B. Giddings, J.A. Harvey, and A. Strominger,
\prd{45} (1992) R1005.}
\REF\sda{S. P. de Alwis, \pl{289}  278 (1992),\hfill\break
 and \pl{300} 330 (1993).}
\REF\bc{A. Bilal and C. Callan, Princeton preprint
PUPT-1320,\hfill\break hepth@xxx/9205089 (1992).}
\REF\shda{S. P. de Alwis, \prd{46} 5429 (1992).}
\REF\rst{J.G. Russo, L. Susskind, L. Thorlacius, \pl{292}, 13 (1992);
L.
Susskind and L. Thorlacius,\np{382}, 123 (1992); J. Russo, L.
Susskind, and L.
Thorlacius, \prd{46}, 3444(1992).}
\REF\stu{L. Susskind, L. Thorlacius, J. Uglum, Stanford preprint,
SU-ITP-93-15,
hep-th/9306069.}
\REF\ct{T. Curtright and C. Thorn, \prl{48} (1982) 1309; E. Braaten,
T.
Curtright, and C. Thorn, \pl{118} (1982) 115; \ap{147} (1983) 365; E.
D'Hoker
and R. Jackiw, \prd{26} (1982) 3517.}
\REF\mass{S. P. de Alwis ``Two Dimensional Quantum Dilation Gravity
and the
Positivity of Energy" Colorado preprint COLO-HEP-309 hepth/9302144.}
\REF\ddf{ See for example, M. Green , J. Schwarz, E. Witten,
``Superstiring
theory:1",  Cambridge (1987).}
\REF\sv{Schoutens, E. Verlinde, H. Verlinde, Princeton preprint,
PUPT-1395,\hfill\break IASSNS-HEP-93/25. See also S. Hirano,
Y.Kazama, and Y.
Satoh, preprint UT-Komaba 93-3.}
\REF\bak{C. Pope, L. Romans, and X.Shen, Phys. Lett. B236 (1990) 173;
I. Bakas,
and  E. Kiritsis, in "Common Trends in Mathematics and Quantum Field
Theories",
Progress of Theoretical Physics, supplement 102 (1990) 15,  edited by
T.
Eguchi, T Inami, and T. Miwa.  }
\REF\raj{See for example, R. Rajaraman, ``Solitons and Instantons: An
Introduction to Solitons and Instantons in Quantum Field Theory",
North-holland, Amsterdam, (1982).}
\pubnum {COLO-HEP-318\cr hepth@xxx/9307140}
\date={July 1993}
\titlepage
\vglue .2in
\centerline{\bf On the Passage from  the Quantum theory to the
Semi-Classical
theory}
\centerline{\bf in 2d Dilaton Gravity }
\author{ S.P. de Alwis\foot{dealwis@gopika.colorado.edu}}
\address{Dept. of Physics, Box 390,\break
University of Colorado,\break Boulder, CO
80309}
\vglue .2in
\centerline{\caps ABSTRACT}
  In this paper an attempt is made to  understand the passage from
the exact
quantum treatment of the CGHS theory  to the semi-classical physics
discussed
by many authors. We find first that to the order of accuracy to which
Hawking
effects are calculated in the theory, it is inconsistent to ignore
correlations
in the dilaton gravity sector. Next the standard Dirac or BRST
procedure for
implementing the constraints is followed. This leads to a set of
physical
states, in which however the semi-classical physics of the theory
seems to be
completely obscured.  As an alternative, we construct a coherent
state
formalism, which is the natural
 framework for understanding the semi-classical calculations, and
argue that it
satisfies all necessary requirements of the theory, provided that
there exist
classical ghost configurations which solve an  infinite set of
equations. If
this is the case it may be interpreted as a spontaneous breakdown of
general
covariance.
  \endpage

The simple two dimensional model for understanding questions
associated with
Hawking radiation [\hawk],  proposed by Callan Giddings Harvey and
Strominger
(CGHS)[\call ], has recently been the basis of much research
activity. In
particular it was pointed out by the present author [\sda], and
independently
by Bilal and Callan [\bc], that general covariance implies that the
quantum
theory has to be a conformal field theory (CFT), and that conformal
invariance
requires that the classical CGHS action be modified by quantum
corrections.
Furthermore it was shown in
[\sda, \bc] that a  class of modifications leads to a solvable
Liouville-like
CFT, and an example of this class was explicitly worked out. Another
example
was given in [\rst]. In both these examples however the
transformations to
the fields of the Liouville-like theory had a singularity  when the
number of
matter fields $N$ was greater than $24$, with the implication that
the
definition
of the exact quantum theory was unclear, since the range of
integration (in
field space) of the
Liouville-like CFT was only half the real line. In a later paper
[\shda] it was
pointed out by the author that there is in fact a class of theories
in which
this problem is absent ( even for $N>24$). Such theories  are
equivalent to an
exact solvable quantum conformal field theory which can be treated by
standard
techniques.

It should be emphasized at this point, that making some modification
of the
original CGHS classical action, in order to get a CFT, is not a
matter of
choice, but a matter of necessity for the consistency of the theory.
While
there can be more general quantum corrections which lead to CFTs
which are not
solvable exactly, it is certainly consistent to focus attention on
the solvable
class. Indeed  at the semi-classical level the required modifications
all lead
to the solvable Liouville-like theory.
Indeed one cannot discuss even the semi-classical physics of the
CGHS theory in a consistent fashion without including these necessary
modifications, since they are of the same order as the anomaly term
introduced
by CGHS [\call]
in order to discuss Hawking radiation.\foot {This point is discussed
in detail
in \shda.}

Now as pointed out in [\mass] it does not seem as if the exact
quantum theories
coming from CGHS are good candidates for discussing Hawking
radiation. In order
 to model the four dimensional situation, it seems necessary
to impose a boundary in the two dimensional space-time corresponding
to the
origin of polar coordinates
 in $3+1$ dimensions. Such a boundary  occurs in the models discussed
in [\sda,
\bc, \rst]. Indeed there is no problem discussing the
semi-classical physics of these theories provided that one imposes a
suitable
boundary condition as  in [\rst]. However the physics one gets out
will
depend on the boundary condition one imposes. More importantly  it is
not at
all clear that this theory defines a consistent quantum theory, since
as we
will
discuss later in some detail, the argument that the Liouville like
theory is
a CFT depends crucially on the assumption that the range of
integration
goes over the whole real line.

The main aim of this paper is  to  understand the passage from the
exact
quantum
treatment to the semi-classical theory. The standard method of
treating a CFT
is to impose the physical state conditions. However there is no trace
of the
black hole in the space of physical states and it is unclear how to
obtain the
semi-classical equations. {\it In particular
 the expectation values of field operators in these states are space
time
independent.} We suggest therefore that this
space should be enlarged by including coherent states  in  which
the expectation values of the (total) stress tensor are zero.  The
independence
 of the  functional integral from the fiducial metric  also requires
that
products of the
stress tensor have vanishing expectation values as well. At short
distances
this is guaranteed by  the zero central charge Virasoro algebra,
which is a
necessary  condition, but is by no means sufficient. To satisfy this
condition
at finite distances it seems that one has to also impose the
condition that all
higher (i.e >2) conformal  spin operators of the thoery also have
zero
expectation value.

An important and related question is the emergence of time in quantum
gravity.
In principle this question should also have a resolution since we
have now an exactly solvable quantum theory.  The physical state
condition is
the precise statement of the Wheeler-DeWit equation (in a Fock basis)
for the
theory. In this case there is no Schrodinger time evolution. It may
also be
noted that the existence of a non-zero boundary  Hamiltonian for a
classical
open system does not make any difference to the evolution of local
operators in
the quantum theory. On the other hand in the coherent state sector
one has the
time
evolution of the expectation values in accord with the correspondence
principle.

Although it was stated already in [\sda, \bc ] that the
Liouville-like theory
is probably an exact CFT (and indeed in [\shda] it was observed that
the
Liouville theory argument of [\ct ] can be extended to our case), a
detailed
argument was not given. In the first part of this paper we will
therefore give
the justification for this statement. Next we discuss the physical
states and
point out that they do not seem to have anything to do with the
semi-classical picture of black hole formation and evaporation. Then
we discuss
our alternative and make some concluding remarks.

Let us first prove our assertion that the Liouville-like theory is a
CFT.
The action is

 $$S={1\over 4\pi}\int d^2\s [\mp\pa_{+}X\pa_{-}
X\pm\pa_{+}Y\pa_{-}Y+\sum_i\pa_{+}f^i\pa_{-}f^i+2\l^2e^{\mp\sqrt{2\ov
er |\k
|}(X\mp Y)}].
\eqn\action$$
 The field variables are related to the original variables $\phi$
(the dilaton)
 and $\rho$ (half the logarithm of the conformal factor) that occur
in the
 CGHS action gauge fixed to the conformal gauge
$(g_{\a\b}=e^{2\rho}\eta_{\a\b})$, by the following relations;
$$Y=\sqrt{2|\k|}[\r-\k^{-1}e^{-2\phi}+{2\over\k}\int d\phi
e^{-2\phi}\ov h
(\phi )].\eqn\ycdt$$
 $$X=2\sqrt{2\over |\k |}\int d\phi P(\phi ),\eqn\xcdt$$
where
$$P(\phi )=e^{-2\phi}[(1+\ov h)^2+\k e^{2\phi}(1+h)]^{1\over
2}.\eqn\pee$$
and $\k ={24-N\over 6}$, $N$ being the number of matter fields. In
\ycdt,
\xcdt, the functions $h(\phi),~\ov h(\phi)$ parametrize quantum
(measure)
corrections that may come in when transforming to the translationally
invariant
measure (see [\sda, \bc, \shda ] for details).  The statement that
the quantum
theory has to be independent of the fiducial metric (set equal to
$\eta$ in the
above) implies that this gauge fixed theory is a CFT. The above
solution to
this condition was obtained by considering only the leading terms of
the beta
function equations, but it was conjectured in [\sda, \bc ] (on the
basis of the
resemblance of \action\ to the Liouville theory), that it is indeed
an exact
solution to the conformal invariance conditions. It should be
stressed  that
this latter statement is strictly valid only when $P$ has no zeroes.
This
implies some restrictions on the possible quantum corrections (as
shown in
[\shda ]  there is a large class which satisfies these conditions).

Putting
$$\z_{+}\equiv \sqrt{2\over |\k |}(X+Y)+\ln{\sqrt{2\over |\k
|}},~~~\z_-\equiv
\sqrt{2\over |\k |}(X-Y),$$

let us define the generating functional for the connected correlation
functions
of the theory by

$$e^{i|\k |W[J]}=\int [d\z_+][d\z_-]e^{i|\k |S[ \z_+,\z_-]+i|\k |\int
d^2\s(J_+\z_-+J_-\z_+)}.\eqn\qft$$
We have ignored the matter ($f$-fields) and the ghost contributions
since they
are  CFTs by themselves.

In terms of the $\z$ fields the action plus source terms take the
form,
$${1\over 4\pi}\int d^2\s
[\pa_+\z_+\pa_-\z_-+2\l^2e^{\z_+}+4\pi(J_+\z_-+J_-\z_+)].\eqn\new$$

The equations of motion are
$$\eqalign{-\pa_+\pa_-\z_-+2\l^2e^{\z_+}+4\pi J_-= & 0\cr
-\pa_+\pa_-\z_++4\pi J_+= & 0.\cr}\eqn\motion$$

The effective action is defined by the following equations.

$$\G [\z_{\pm c}]=W[J]-\int d^2\s (J_+\z_{-c}+J_-\z_{+c}),\eqn\eff$$
$$\z_{\pm c}={\d W\over\d J_{\mp}}=<\z_{\pm}>_J\equiv{\int
[d\z_{\pm}]\z_{\pm}e^{iS+iJ.\z}\over\int[d\z_{\pm}e^{iS+iJ.\z}}\eqn\c
lfield$$

It follows that
$${\d\G\over\d \z_{\pm c}}=-J_{\mp}\eqn\jeqn$$

Since the measures $[d\z_{\pm}]$ are translationally invariant we may
use the
equations of motion inside the functional integral to get
$$\pa_+\pa_-{\d W\over\d J_-}=\pa_+\pa_-\z_{+c}=4\pi J_+,\eqn\zplus$$
$$\pa_+\pa_-{\d W\over\d J_+}=\pa_+\pa_-\z_{-c}=2\l^2<e^{\z_+}>+4\pi
J_-.\eqn\zminus$$

{}From \zplus\ we have
$$W[J]=\int d^2\s d^2\s 'J_-(\s )\D_F(\s -\s ')J_+(\s )+\int d^2\s
\z_{+c}^0J_-+F[J_+],\eqn\wj$$
where the last term is a functional (to be determined below) of $J_+$
only, and
$\z^0_{+c}$ is a  c-number solution of the homogeneous equation.
$\D_F$ is the
Feynman-Green function defined as the solution of
$\pa_+\pa_-\D_F(\s-\s ')=4\pi\d^2 (\s-\s ')$ with the Feynman
boundary
conditions which come from the fact that the path integral actually
must
contain the initial  and final  ($t\rightarrow\pm\infty$) wave
functionals.
{}From equation \wj\ for $W$ we  have, for the connected correlation
functions,
$$<\z_+(\s )\z_+(\s ')>_{conn,J}=(i|\k |)^{-1}{\d^2W[J]\over\d J_-(\s
)\d
J_-(\s ')}=0,\eqn\corplus$$
and

$$<\z_+(\s )\z_-(\s ')>_{conn,J}=(i|\k |)^{-1}{\d^2W[J]\over\d J_-(\s
)\d
J_+(\s ')}=(i|\k |)^{-1}\D_F (\s -\s ').\eqn\corplmin$$

{}From \corplus\ we see that $<\z_+(\s_1)\ldots
\z_+(\s_n)>=\z_{+c}(\s_1)\ldots
\z_{+c}(\s_n)$ and hence we may rewrite \zminus\ as
$$\pa_+\pa_-\z_{-c}=2\l^2e^{\z_{+c}}+4\pi  J_-.\eqn\zminusnew$$

{}From \zplus, \zminusnew, and \jeqn, we find the effective action,
$$\G (\z_{+c},\z_{-c})={1\over 4\pi}\int d^2\s
[\pa_+\z_{+c}\pa_-\z_{-c}+2\l^2e^{\z_{+c}}].\eqn\neweff$$

This effective action has already been derived by Russo, Susskind,
and
Thorlacius [\rst], by formally doing the $\z_-$ integral in \qft\
which gives a
delta functional that allows one to do  the $\z_+$ integral. We have
however
obtained this by a long winded route because it is important for our
purposes
to demonstrate exactly and rigorously what the correlation functions
are. From
the effective action we obtain, by doing the inverse Legendre
transform, the
generating functional for connected correlation functions as \wj,
with

$$ F[J]={\l^2\over 2\pi}\int d^2\s e^{[\int d^2\s '\D_F(\s-\s
')J_+(\s
')+\z_{+c}^0]}.$$

Using this we have the final
correlation function
$$<\z_-(\s)\z_-(\s ')>_{conn,J}={\l^2\over 2\pi}{2\over |\k |}\int
d^2\s ''\D
(\s -\s '')e^{<\z_+(\s'')>_J}\D (\s ''-\s ').$$

 It should be stressed however that the above arguments  depend
crucially on
using the equations of motion inside the functional integral. i.e. we
have used
the formula
$$\int [d\z]{\d\over\d \z(\s )}e^{iS+iJ.\z}=0,$$
which will not be valid unless the functional integration ranges over
the whole
real line. This in turn means that the argument is valid only for
quantum
theories in which $P(\phi )$ has no zeroes so that the integration
range for
$\z$ is not cut off. Now we are not claiming that the theories in
which there
is a so-called quantum singularity are necessarily inconsistent, but
only that
it is very difficult to establish that they are a correct
representation of the
quantum theory, in so far as it is not at all clear that they are
CFTs. The
present author's hunch is that this singularity is spurious. That
this may be
the case is also acknowledged in [\rst ].

It follows also that there is an operator formulation of the theory
with
operator equations of motion

$$\pa_+\pa_-\hat \z_+=0,~~\pa_+\pa_-\hat \z_-=2\l^2e^{\hat
\z_+}\eqn\eofm$$

and equal time commutation relations,

$$\eqalign{
[\hat \z_+(\s ),\dot{\hat \z_-}(\s ')]\d (t-t') & ={8\pi i\over |\k
|}\d^2
(\s-\s ')\cr
[\hat \z_-(\s ),\dot{\hat \z_+}(\s ')]\d (t-t') & ={8\pi i\over |\k
|}\d^2
 (\s-\s ')\cr}\eqn\comm
 $$
with all other commutators vanishing.
One could  have arrived at this directly by means of a canonical
quantization of the action \action\ with the conjugate momenta
$\Pi_{\pm}={|\k
|\over 8\pi}\dot \z_{\mp}$. {\it  However the path integral
formulation
highlights the fact that these operator equations are valid only for
those
theories for which the field has a range which extends
over the whole real line.} From now on we will deal with the operator
formulation  and will  drop the hats on operators.

The stress tensors may be expressed in terms of the canonical
variables using
the equations of motion to rewrite the second order time derivative
terms, and
we get

$$\eqalign{T_{\pm\pm}=&{|\k |\over
4}(:\pa_{\pm}\z_+\pa_{\pm}\z_-:+\pa_{\pm}^2(\z_+-\z_-))\cr
=&T^0_{\pm\pm}-{|\k
|\over 2}\l^2e^{\z_+},\cr}$$

where $T^0$ is the stress tensor of the free $(\l^2=0)$ theory
expressed in
terms of the canonical variables. Then it easily follows from the
fact that the
exponential term is a dimension $(1,1)$ operator with respect to
$T^0$ [\sda,
\bc ], which has no correlations with itself \corplus, that
the full stress tensor generates a Virasoro algebra with central
charge
$c_\z=2+6\k$.  In addition it follows from the equations of motion
that

$$T_{+-}=0,~~\pa_{\mp}T_{\pm\pm}=0,$$

as operator statements.
Thus we have the complete operator formulation of a CFT. The matter
and ghost
sectors (in conformal gauge) are free  CFTs with central charges
$c_f=N,~c_{gh}=-26$, so that the quantum CGHS theory is a CFT
with zero central charge for $\k={24-N\over 6}$.

For future reference we also note the following.
The solutions to the equations of motion are

$$\eqalign{\z_+(\s,\tau )=&g_+(\s^+)+g_-(\s^-)\cr
\z_-(\s,\tau
)=&u_+(\s^+)+u_-(\s^-)+2\l^2\chi_+(\s^+)\chi_-(\s^-)\cr}\eqn\soln$$

where, $\chi_{\pm}(\s^{\pm})=\int^{\s^{\pm}}e^{g_{\pm}}$,  and
$\s^{\pm}=\tau\pm\s$.

Also from the equal-time commutators and these solutions one  obtains
the
following light-cone commutator,

$$[\pa_{\pm}\z_-(\s^{\pm}),\chi_{\pm}(\s'^{\pm})]={8\pi i\over |\k
|}\t
(\s^{\pm}-\s'^{\pm})\pa_{\pm}\chi_{\pm}(\s )\eqn\chicomm$$

Now the general covariance of the theory is reflected in the
conformal gauge by
the constraints, which  classically amount to the statement that the
total
stress tensor must vanish. In the quantum theory general covariance
has the
implication that the gauge fixed path integral be independent of the
fiducial
metric $\hat g$ [\sda,\shda ].\foot{ The physical metric, which is
integrated
over, can always be put in the form $g=e^{2\rho}\hat g$ in two
dimensions.} In
the semi-classical analyses of the theory [\call, \sda, \bc, \shda,
\rst, \stu,
etc.] this was only implemented in the form

$$<\Psi |T_{\pm\pm}|\Psi >=0\eqn\semicons$$

 where,

 $$T=T^{dg}+T^f+T^{gh}$$

 is the total stress tensor of dilaton gravity $dg$, matter $f$, and
diffeomorphism ghosts $gh$. Once the fiducial metric is fixed  this
constraint
gives a relation between (the expectation values of)
 matter field and the dilaton gravity fields.\foot{We may set the
expectation
value of the ghost stress tensor to zero in a particular conformal
frame,
 for example the Kruskal one in which $\rho=\phi$, see [\shda ].} One
then
discusses Hawking radiation by transforming to coordinates
appropriate to an
asymptotic observer. As pointed out in  [\shda, \mass ] this
procedure already
has some ambiguites stemming from the fact that it is only the total
stress
tensor which transforms as a tensor, and that it is not possible to
give a
coordinate invariant justification for the usual argument. The
situation
becomes even more confusing if one follows the standard prescription
for  an
exact quantum treatment. The reason is that the  although equation
\semicons\
is a
 necessary condition it is  by no means a sufficient expression of
the
constraints of coming from general covariance. In fact the complete
statement
of the constraints in the quantum theory (i.e. the independence of
the quantum
theory from the fiducial metric) is that the physical states of the
theory must
be such that the expectation value of an arbitrary product of stress
 tensors must vanish.

  $$<\Psi|\prod^n_{r=1} T_{\pm\pm}(\s_r)|\Psi>_{conn}=0\eqn\cons$$

 Since we have a Virasoro Algebra with zero central charge the
leading
singularities of \cons\ are  guaranteed to vanish by \semicons. But
in general
the physical states of the quantum theory belong to a smaller
subspace than
those allowed by \semicons. In fact it is possible to argue that it
is
inconsistent to ignore the $n=2$ constraint in \cons\ when discussing
Hawking
radiation, which is an $O(\hbar )$ effect in \semicons. To see this
let us note
that we may introduce the analog of the Einstein tensor $G_E$ in $4d$
by
writing $T^{dg}=NG_E$, so that \semicons\ takes the form
$N<G^{dg}>=-<T^f>$.\foot{We  ignore the ghosts at this point, and
note that
$N$ plays the role of the inverse Newton constant here.} Hawking
radiation is
an $O(N\hbar )$ contribution to the
right hand side which must be balanced by an $O(\hbar )$ contribution
to $G_E$.
Now the $n=2$ equation in \cons\ may be written
$N^2<G_EG_E>=<T^fT^f>$. The
correlations of the $N$ matter fields give $O(N\hbar )$ contribution
on the
right hand side which can only be balanced by the $O(N^{-1}\hbar )$
correlations (see \corplmin) of the dilaton gravity fields on the
left hand
side. In other words the constraints inform us that to the same order
to which
we are calculating the back reaction to Hawking radiation in
\semicons\ we must
also take into account the $n=2$ equation in \cons.

The standard way of ensuring that \cons\ is satisfied is to implement
the
physical state condition i.e.,

$$T_{\pm\pm}^+|\Psi >=0,~ or~ Q_B|\Psi >=0,\eqn\phy$$

where $T^+$ is the positive frequency part of the stress tensor and
$Q_B$ is
the BRST charge. Using standard string theory methods it is possible
to obtain
the solution to these conditions. One first constructs an operator
(essentially
the so-called DDF operator of string theory [\ddf ] ) by dressing the
matter
field
with the dilaton gravity field $\chi$, which it should be noted (see
(20) and
(14)) has no correlations with itself.

$$A^i_{\pm}(\o )={1\over 2\pi}\int
d\s^{\pm}(\l\chi_{\pm})^{i\o}\pa_{\pm}f^i\eqn\ddf$$

It is easily shown (using (21)) that these operators commute with the
total
stress tensor. \foot{For the case $N=24$ this result has already been
obtained
in [\sv ]}. Furthermore they satisfy the same commutation relations
as the mode
operators of the $f$ field. Thus the action of products of these
operators
(with negative frequencies) on the Fock vacuum $|0>~~ (T^+|0>=0)$
give
solutions to the physical state conditions. i.e.

$$|\Psi >=\prod A|0>.\eqn\phystate$$

It is probably  the case that the standard argument that these states
are
complete, in the space of physical states up to spurious (i.e. BRST
trivial )
states, is also valid here modulo technicalities involving the
infinite length
of the space , but we will not discuss this further,  since it is
not germane
to the main point of this paper.\foot{  A detailed argument to this
effect  has
been given for the case $N=24$, with reflecting boundary conditions
and an
infra-red cut off, in [\sv].}

The question we wish to address is the interpretation of the
semi-classical
equations that have been discussed in the literature. As long as one
did not
have an exact solution or even a formulation of quantum gravity,  one
is free
to speculate as to what it might be. However  the model that we have
is a
well-defined and solvable theory of quantum gravity, albeit in two
dimensional
space time. Nevertheless as we will
see, if we impose the standard rules for this theory (i.e. \phy ), it
is very
difficult to extract the semi-classical equations which have been
used to
discuss  issues like
Hawking radiation.

We begin with an elementary observation. The physical state condition
\phy\
implies in particular that

$$H|\Psi >=0,~~P|\Psi >=0,\eqn\nospacetime$$

where the Hamiltonian and momentum operators are given by

$$H=\int d\s^+ T_{++}+\int d\s^- T_{\--},~ P=\int d\s^+ T_{++}+\int
d\s^-
T_{\--}.$$

Now the semi-classical theory deals with space time dependent
classical fields
(for both dilaton gravity and matter), which are  to be interpreted
as the
expectation values of field operators in some quantum state. Clearly
such a
state cannot be one satisfying \phy\ since from \nospacetime\
$<\Psi|\hat\phi
|\Psi >$ is independent of space-time. From our explicit construction
it is
also  clear that there is no limit in which we can recover any thing
like a
semi-classical state.  This is in contrast to the usual argument that
there
must be some limit, $G_N<<1$ in four dimensions, $N\rightarrow\infty
$ in the
CGHS case, in which quantum gravity should yield a semi-classical
formulation.
Instead what we have  established is that  the principal difference
between the
semi-classical theory and the usual formulation of the quantum theory
lies in
the way the constraints are imposed. In the former case one imposes
the
constraint as an expectation value \semicons\ in the latter as a
condition on
the states \phy.

Is there an alternative to imposing the physical state conditions
\phy ? Since
the physical requirement is actually \cons, and since \phy, though a
sufficient
condition, has not been shown to be necessary, one might ask whether
it is too
strong a condition. In other words the question is whether there are
states
which satisfy \cons\ exactly without  satisfying \phy . This is
tantamount to
asking whether general covariance can be broken
spontaneously.\foot{The Hilbert
space of the theory has negative norm states and therefore the usual
injunction
against spontaneous symmetry breaking in $2d$ will not apply for the
same
reason that there is no Goldstone theorem in a gauge theory. Of
course in $2d$
there are no gravitons either!} We believe that the following is
a possible solution to this problem.

The natural quantum states for representing classical configurations
are
coherent states. In order to understand the essence of the problem it
is
helpful to consider first the example of the harmonic oscillator.
Introducing
creation and annihilation operators $a, a^{\dag},$ satisfying the
comutation
relation $[a,a^{\dag}]=1$, a (normalized) coherent state of the
oscillator is
given by $|z>=e^{-{z\ov z\over 2}}e^{z a^{\dag}}|0 >$. This has the
property
that
$a|z>=z|z>,$ so that in particular the position operator has the
expectation
value $\Re z$ and $<z|:h(a^{\dag}, a):|z>=h(\ov z, z)$. The analog of
the
physical state constraint \phy\  would be a requirement that only
(say) the
unit energy eigenstate\foot{The zero energy state is  trivially equal
to the
$z=0$ coherent state so a non-zero energy state gives a better
analogy.}
 $|1 >$ of the Hamiltonian $ a^{\dag}a$ is ``physical". This state is
ofcourse
an infinite superposition of the coherent states,
$$|1 >=\int dzd\ov z|z><z|1 >.\eqn\cohsup$$

Now semi-classical physics in our case picks up (the analog of) one
state from
this infinite superposition so although the ``physical state" itself
will not
represent space time dependent configurations each member of the
superposition
separately may. Indeed since $|<z'|z>|=e^{-{(|z|^2+|z'|^2)\over
2}}e^{\Re (\ov
z'z)}\rightarrow 0$ for large $|z|$ one may argue that  large  field
configurations will exhibit classical behaviour and one will not be
able to
detect interference with other members of the superposition \cohsup.
It should
be noted that  according to the ``physical state" condition the
system must be
in $|1>$ which is not a complete set of states for the whole Hilbert
space. In
particular a coherent state cannot be expressed as a superposition of
physical
states.

Let us now get back to dilaton gravity.
The constraint equations of the semi-classical theory are easily
obtained as
exact statements by taking $|\Psi >$ in  to be a coherent state in
\semicons.
In the following  $u$ and $g$ are classical $c$-number  fields such
that the
classical solutions to the equations of motion $\z_{cl}=<\z >$ are
given in
terms of them by \soln. Working in the Kruskal coordinate system
where
$\z_{cl}=g=0$,  we take

$$|\Psi >=NV_fV_{\z}V_{gh}|0 >\eqn\coh$$

where $N$ is a normalization factor and,

$$\eqalign {V_f & =  e^{{1\over 8\pi i}\int d\s f^{(-)}\pa
f_{cl}^{(+)}}\cr
V_{\z} &  = e^{{1\over 8\pi i}\int d\s \z^{(-)}_+\pa u^{(+)}}\cr}$$

 with a similar expression for the ghost sector involving classical
ghost
configurations (Grassmann valued functions) $b_{cl},~c_{cl}$.
\foot{The
superscript $(-)$ is an instruction to take the negative frequency
part of the
corresponding field.} These states do not satisfy the physical state
condition
\phy. But, as we discussed in our harmonic oscillator example any
physical
state can be expressed as a superposition of these coherent states.
For large
field configurations one may argue again that interference with other
members
of the superposition will not be detectable.
 The semi-classical analysis however also requires that \semicons\ be
imposed.
 If one is starting with a physical state however, this equation
(actually all
of the equations in \cons) are satisfied automatically. There is no
reason why
any given member of the superposition should be made to satisfy
\semicons\
separately.  Indeed the physical  state condition would be satisfied
through
cancelations
 between different members of the superposition of coherent states.

 To make sense of the semi-classical equations then one has to try to
satisfy
 \cons\ on coherent states.
For $n=1$ i.e.  \semicons, one has  the  equations used in the
semi-classical
analyses, [\call, to \stu]

 $$\pa_{\pm}^2u +\pa_{\pm}f_{cl}\pa_{\pm}f_{cl}+t_{\pm\pm}=0$$

 where $t$ is the classical ghost stress tensor.
  Now while this is an exact equation in the quantum theory (because
of our use
of coherent states) the statement of general covariance in the
quantum theory
is by no means exhausted by this.
  One has to also satisfy the
  infinite set of equations \cons. In the semi-classical discussion
  only the first of these was satisfied. When $n>1$ in \cons\ the
leading
singularities in the correlation functions are also guaranteed to
vanish by
  the Virasoro algebra with zero central charge and \semicons.
However the
subleading terms (including the non-singular ones)   of the operator
product
expansion for products of stress tensors gives an infinite number of
constraints of the form $ <W_n>=0$,
 where $W_n$ are  the generators of the  $W_{\infty}$ algebra of the
system. It
should be noted that since the central charge of the Virasoro algebra
is zero
so is the central charge of the $W$ algebra so that there is no
c-number term
coming from multiple $(n>2)$ operator products [\bak ].
 These additional constraints imply  that the classical ghost
configurations in
our coherent state satisfies an infinite set of conditions. It is not
clear
that a solution exists, but on the other hand there is no obvious
reason why
this is ruled out either. If such a solution exists then we would
have achieved
a spontaneous breakdown of general covariance.

  The CGHS theories in which the field space is unrestricted, are
exact
solvable quantum CFTs. Therefore although they are not good models
for
understanding Hawking radiation in $3+1$ dimensions, in so far as
they are
precisely defined theories of two dimensional quantum gravity,
they may be used to elucidate the conceptual problems associated with
quantization of geometry. In fact what we have established is that in
our
simple theory, the so-called semi-classical equations have nothing to
do with
taking a large $N$
limit.\foot{Ofcourse if the theory with a boundary [\rst] is a
consistent
quantum theory, i.e. is a CFT, then it is likely to have very
complicated
higher order effects which would be suppressed at
$N\rightarrow\infty$.
Nonethelss insofar as it must be a CFT, the basic problem of how to
impose the
constraints, will be exactly as in the simple theory discussed in
this paper.}
In the standard treatment of the quantum theory
using Dirac quantization or BRST techniques the semi-classical
picture of black
hole formation and decay seems to be completely hidden from view.
Presumably
this is because one is considering quantum states which are generally
covariant
and are a superposition of all metrics.\foot{ In this sense the state
is
similar to one in the soliton sector in a quantum treatment of field
theories
with soliton solutions, where the underlying translational symmetry
of the
theory is restored by integrating over  collective coordinates. [\raj
]} In
this picture then one needs some mechanism for collapsing the
geometrical part
of the wave function to one metric configuration.

  An alternative to the standard picture has been suggested here by
weakening
the constraint on the states (which is a sufficient but not a
necessary
condition for the general covariance of expectation values). This
leads to a
formalism using coherent states which gives the usual semi-classical
equations,
and indeed in our simple theory the latter would be exact. However
it is not
clear that the  formalism is completely consistent in that we are
unable to prove that the infinite set of equations for the ghost
configurations
can be satisfied.

The original motivation for studying the CGHS theory was to provide a
simple
toy model in which the question of information loss and the breakdown
of
unitary evolution could be decided one way or the other. All
the arguments on either side of this issue even in this simplified
context have
been  semi-classical with the differences between the two (or three)
camps
depending on the assumptions made about the quantum gravity regime.
Indeed in a
recent paper Susskind, Thorlacius, and Uglum [\stu ] have argued that
the
separation of scales  that is usually used to justify semi-classical
physics is
not valid in the analysis of Hawking radiation. Furthermore they
propose a
principle of complementarity between the physics obtained by an
asymptotic
observer and that of one falling into the black hole.

What we have proposed here is that one should try to test such
hypothesis
within the well defined context of quantum dilaton gravity. Indeed
given that
there is no experimental test which can decide these issues it is
imperative
that the logical basis of the arguments used are as firm as possible.
In
particular since underlying reality is quantum mechanical, and
geometry itself
must be quantum mechanical, the logical problem becomes one of
deriving the
observed classical (or semi-classical) world from the quantum
mechanical one.
As we have seen, it is very difficult to do this even in
our simple theory. This is probably  an indication
that classical (or semi-classical) intuition may break down in
unexpected ways.

{\bf Acknowledgements:} I would like to thank , T. Eguchi, G.
Horowitz,  J.
Polchinski, A. Strominger, L. Susskind, and especially R. Brustein,
for
discussions. Part of this work was done while the author was
participating in
the work shop ``Non-perturbative methods in string theory" at the
Institute for
Theoretical Physics at UC Santa Barbara, and this research was
supported in
part by the National Science Foundation under Grant No. PHY89-04035.
This work
is also partially supported by Department of Energy contract No.
DE-FG02-91-ER-40672.

\refout
\end